\documentclass[12pt]{iopart}

\usepackage{graphicx}
\usepackage{multicol}

\begin{document}

\title[Quantum enhancement of $N$-photon phase sensitivity]{Quantum enhancement of $N$-photon phase sensitivity by interferometric addition
of down-converted photon pairs to weak coherent light}

\author{Takafumi Ono and Holger F. Hofmann}
\address{
Graduate School of Advanced Sciences of Matter, Hiroshima
University, Kagamiyama 1-3-1, Higashi Hiroshima 739-8530, Japan}

\begin{abstract}
It is shown that the addition of down-converted photon pairs to
coherent laser light enhances the $N$-photon phase sensitivity due to the
quantum interference between components of the same total photon
number.
Since most of the photons originate from the coherent laser
light, this method of obtaining non-classical $N$-photon states is
much more efficient than methods based entirely on parametrically
down-converted photons. Specifically, it is possible to achieve
an optimal phase sensitivity of about $\delta \phi^2=1/N^{3/2}$, equal to
the geometric mean of the standard quantum limit and the Heisenberg
limit, when the average number of down-converted photons
contributing to the $N$-photon state approaches
$\sqrt{N/2}$.
\end{abstract}

\pacs{
42.50.Dv 
42.50.St 
03.67.-a 
42.50.Lc 
}

\ead{h.hofmann@osa.org}


\maketitle

\section{Introduction}
Recent advances in the generation and control of non-classical
multi-photon states have made it possible to realize
super-sensitive phase measurements, where quantum correlations
between the photons reduce the errors in a phase measurement
below the standard quantum limit valid for uncorrelated photons
\cite{Gio06}. Up to now, many of the theoretical and
experimental efforts have focused on the generation of
path entangled states (also known as NOON states), where
the photons are in a superposition of all photons in one path
or all photons in the other path \cite{Eda02,Kok02,Fiu02,Zou02,Pry03,
Hof04,Wal04,Mit04,Sha04,Eis05,Liu06,Nie07,Cab07}. 
However, experimental noise and
low post-selection efficiencies have made it difficult to overcome
the standard quantum limit at photon numbers higher than two
\cite{Res07}. It is therefore useful to consider alternative
approaches that do not require maximal path entanglemed input
states \cite{Hol93,Ste02,Wan05,Hof06,Sun06,Nag07,Sun07}.
In particular,
these proposals make more direct use of the standard source of
non-classical light, the emission of photon pairs by spontaneous
parametric down-conversion. It might also be worth noting that the
first experiment to actually beat the standard quantum limit
with a four photon state \cite{Nag07} was based on one of these
proposals \cite{Ste02} and used the four photon component
$\mid 2;2 \rangle$ of two mode down-conversion as input state.

Yet another alternative approach towards achieving optical
interferometry beating the standard quantum limit is the use
of squeezed states \cite{Cav81,Yur86,Kit93,Hil93,Win94,Par95,Ber00,Com05}.
These states are usually characterized by a continuous
parameter describing the gradual suppression of quantum noise
observed in the output of the interferometer.
Squeezing can thus describe the transition from weak to strong
non-classical effects within the same conceptual framework.
It has already been known for a long time that improved phase
sensitivity can be achieved by using a squeezed vacuum input
in the empty port of a two path interferometer \cite{Cav81}.
However, the improvement of phase sensitivity is then only equal
to the quadrature squeezing in the input. The recent breakthroughs
in phase sensitive measurements based on path entanglement
show that such high levels of quadrature squeezing are not
necessary to achieve greatly improved phase sensitivities. Instead,
it is sufficient to use spontaneous parametric down-conversion,
which is formally equivalent to a squeezed vacuum with extremely
low squeezing levels. The improvement of phase sensitivity then
originates from multi-photon quantum interferences in the
$N$-photon component detected in the output.
It may thus be possible to achieve high levels of phase
squeezing in $N$-photon states by using multi-photon quantum
interference effects instead of quadrature squeezing, in closer
analogy to the methods employed to generate maximally path
entangled states.

In this paper, we investigate the possibility of gradually
squeezing the $N$-photon component of a weak coherent light input
from a single mode laser by interferometrically adding photon
pairs from a much weaker parametric down-conversion. It is
shown that the quantum interference between the generation of
photon pairs in the laser and the generation of down-converted
photon pairs results in $N$-photon squeezing, where the
squeezing parameter is given by $N$ times the ratio of the pair
generation amplitude $\gamma$ and the squared amplitude of the
coherent light $\alpha^2$. Since $\gamma$ is the
quadrature squeezing parameter of the down-converted
light, the selection of the $N$-photon component in
coincidence counting thus amplifies the squeezing by a
factor of $N/\alpha^2$. It is therefore possible to
increase the squeezing level obtained by reducing the
amplitude of the coherent input light. We find that
this kind of squeezing can
achieve a maximal phase sensitivity of about $\delta \phi^2=1/N^{3/2}$ when the average number of
down-converted photons contributing to the $N$-photon state
approaches $\sqrt{N/2}$.
Even at this optimal squeezing value, most of the photons
originate from the coherent laser light. Therefore, high photon
numbers $N$ can be obtained even without particularly bright down-conversion sources. The interferometric addition of down
converted photon pairs to coherent laser light thus provides an
extremely efficient tool for beating the standard quantum limit
at high photon numbers.

The rest of the paper is organized as follows.
In section \ref{sec:MZ} we review the quantum mechanics of
optical phase estimation and derive a definition of phase
squeezing based on the Hilbert space representation of
uncertainties.
In section \ref{sec:qi}, we show how destructive interference
between the uncertainties of two quantum state components
can be used to achieve this kind of phase squeezing.
In section \ref{sec:operators}, we use operator relations to
derive the general squeezing characteristics obtained by
quantum interference between laser light and down-converted
photon pairs. It is shown that the amount of squeezing
in the $N$-photon component depends on a single parmeter,
$\eta=N \gamma/\alpha^2$. Significant amounts of squeezing
can therefore be obtained if $N \gg \alpha^2$.
In section \ref{sec:approx}, the limits of interferometric
squeezing are considered. It is shown that this kind
of squeezing can
reduce the phase error to $\delta \phi^2 =1/N^{3/2}$, the
geometric mean of the standard quantum limit and the Heisenberg
limit. In section \ref{sec:state}, we describe the squeezed state
in the input photon number basis and analyze the efficiency of
the $N$-photon state generation. It is shown that the probability
of generating an $N$-photon state is orders of magnitude higher
than a corresponding pair state generation with two mode
down-conversion. In section \ref{sec:numerics}, we illustrate the
effects of squeezing by presenting numerical results for the
eight photon case. The relation between classical field interference
and quantum interference effects is illustrated by graphs showing
the photon number statistics of the eight photon interference fringe. In section \ref{sec:conclusions}, the results are summarized and
conclusions are presented.

\section{Phase measurement and squeezing}
\label{sec:MZ}
\begin{figure}[b]
\begin{center}
\includegraphics[scale=0.8]{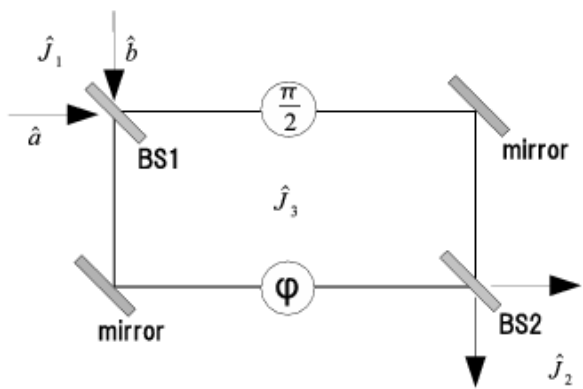}
\caption{\label{MZ}
Illustration of the quantum mechanics of interference in
a Mach-Zehnder interferometer. Light enters the interferometer
in the input modes $\hat{a}$ and $\hat{b}$. The phase difference
between the two paths inside the interferometer is $\phi-\pi/2$.
The photon statistics can be described by a three dimensional
vector $\hat{\bf J}$, where $J_1$, $J_2$ and $J_3$ correspond
to half of the photon number differences between the input modes,
between the two paths inside the interferometer, and between the
output modes at $\phi=0$, respectively.}
\end{center}
\end{figure}

To illustrate the quantum mechanics of N-photon interference, it
is useful to consider a conventional Mach-Zehnder interferometer,
as shown in fig. \ref{MZ}. The light enters the interferometer
in the input modes $a$ and $b$ described by the corresponding
annihilation operators $\hat{a}$ and $\hat{b}$. The two modes then
mix at the input beam splitter BS1, so that each path inside the
interferometer is represented by an equal superposition of
$\hat{a}$ and $\hat{b}$.
Inside the interferometer, each path experiences a different phase
shift. However, only the phase difference between the two paths is
relevant for the observed interference. In the following, we assume
that the phase difference is $\phi-\pi/2$, so that, at $\phi=0$, the
output modes obtained after the path modes interfer at the output
beam splitter BS2 are also equal superpositions of $\hat{a}$ and
$\hat{b}$.

\begin{figure}[t]
\begin{center}
\includegraphics[scale=1.0]{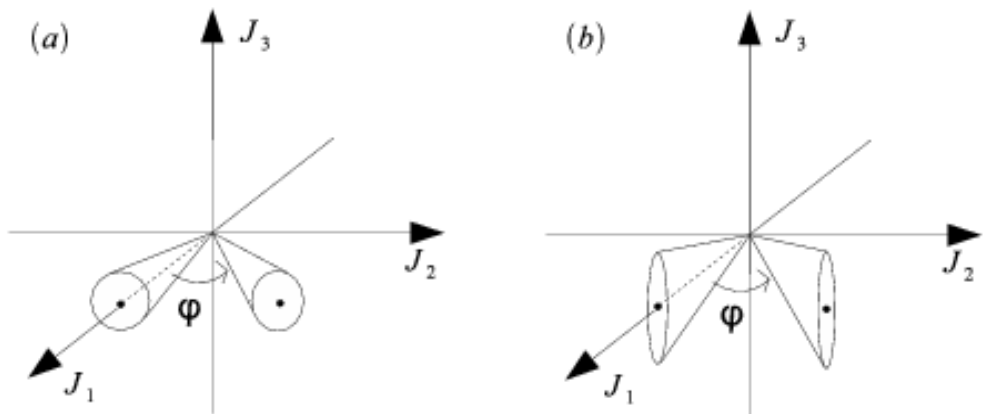}
\caption{\label{SU2}
Illustration the effects of phase shifts $\phi$ on quantum
states using the vector $\hat{\bf J}$ of the Schwinger
representation.
(a) shows the $\mid N;0 \rangle$ state, generated by laser light
in mode $\hat{a}$ and vacuum in mode $\hat{b}$.
Since the $\hat{J}_1$ eigenvalue is maximal, the state is
pointing in the $\hat{J}_1$ direction, surrounded by quantum noise
equally distributed between $\hat{J}_2$ and $\hat{J}_3$.
(b) shows an $N$-photon squeezed state with reduced fluctuations in
$\hat{J}_2$ and correspondingly increased fluctuations
in $\hat{J}_3$.
}
\end{center}
\end{figure}

As indicated in fig.\ref{MZ}, we can now introduce the Schwinger
representation of the two mode photon statistics by
identifying $\hat{J}_1$ with half the photon number difference
between the input modes, $\hat{J}_2$ with half the photon number
difference
between the output modes at $\phi=0$, and $\hat{J}_3$ with
half the photon number difference between the two paths inside
the interferometer. In terms of the input modes $\hat{a}$
and $\hat{b}$, the operators of the Schwinger representation read
\begin{eqnarray}
\label{eq2:schwinger}
\hat{J}_1 &=&
\frac{1}{2}(\hat{a}^{\dagger}\hat{a}-\hat{b}^{\dagger}\hat{b})
\nonumber\\
\hat{J}_2 &=&
\frac{1}{2}(\hat{a}^{\dagger}\hat{b}+\hat{a}\hat{b}^{\dagger})
\nonumber\\
\hat{J}_3 &=&
-\frac{i}{2}(\hat{a}^{\dagger}\hat{b}-\hat{a}\hat{b}^{\dagger}).
\end{eqnarray}
For an $N$-photon state, the mathematical properties of these
operators are identical to the spin operators for a spin quantum
number of $j=N/2$.
It is thus possible to illustrate the possibility of enhancing
the phase sensitivity
geometrically by a three dimensional vector $\hat{\bf J}$, as shown
in fig. \ref{SU2}.
In particular, it is possible to represent
the phase shift $\phi$
in the interferometer as a rotation of the vector $\hat{\bf J}$
around the $\hat{J}_3$ axis. The output photon statistics can then
be expressed in terms of the $\hat{\bf J}$-statistics of the
input state by transforming the output operator $\hat{J}_2$, so that
\begin{eqnarray}
\label{eq2:Jphi}
\hat{J}_2(\phi) &=& \exp\left(i \phi \hat{J}_3\right)
\hat{J}_2 \exp\left(-i \phi \hat{J}_3\right)
\nonumber\\ &=& \cos(\phi) \hat{J}_2(0)  + \sin(\phi) \hat{J}_1(0).
\end{eqnarray}
This relation expresses the complete dependence of the
observable $\hat{J}_2$ measured in the output of the Mach-Zehnder
interferometer on the phase $\phi$.

The dependence of the output statistics of $\hat{J}_2$ on the
phase shift $\phi$ can be used to estimate the value of $\phi$
from the measurement results. For small phase shifts, the
change of $\hat{J}_2$ is proportional to $\phi$, so a
particularly intuitive phase estimate is obtained by dividing
the measurement result of $\hat{J}_2$ by the phase derivative
of the expectation value of $\hat{J}_2$. The sensitivity of this
estimation procedure is limited by the uncertainty of the
estimator observable $\hat{J}_2$ in the input of the interferometer.
Specifically, the phase error $\delta \phi^2$ is
given by the ratio of the squared $\hat{J}_2$-uncertainty
$\Delta J_2^2$ and the squared phase derivative of the
expectation value of $\hat{J}_2$,
\begin{equation}
\label{eq2:deltaphi}
\delta \phi^2 = \frac{\Delta J_2^2}{|\frac{\partial}{\partial \phi}
\langle \hat{J}_2 \rangle|^2} =
\frac{\Delta J_2^2}{\langle \hat{J}_1 \rangle^2}.
\end{equation}
The phase sensitivity of the input state can thus be improved
by reducing the uncertainty in $\hat{J}_2$ while maintaining a
high expectation value $\langle \hat{J}_1 \rangle$.

To determine the kind of quantum states that are particularly
suitable for the phase estimation procedure described above, it
is useful to take a closer look at the Hilbert space geometry
describing the uncertainty limit to the phase sensitivity given
by eq.(\ref{eq2:deltaphi}). Assuming that
$\langle \hat{J}_2 \rangle =0 $,
the squared $\hat{J}_2$-uncertainty of an arbitrary input state
$\mid \psi \rangle$ can be  written as
\begin{equation}
\label{eq2:deltaJ2}
\Delta J_2^2 = \langle \psi \mid \hat{J}_2^2 \mid \psi \rangle
= \langle \psi \mid \! \hat{J}_2 ~\hat{J}_2 \! \mid \psi \rangle.
\end{equation}
Therefore the squared uncertainty $\Delta J_2^2$ is the
inner product of the uncertainty vector
$\hat{J}_2 \mid \! \psi \rangle$ with itself and the
$J_2$-uncertainty $\Delta J_2$ is equal to the
length $||\hat{J}_2 \mid \! \psi \rangle ||$ of this Hilbert space
vector. On the other hand, the phase
derivative of the expectation value of $\hat{J}_2$ can be
expressed in terms of the commutation relation of the
generator $\hat{J}_3$ and the estimator $\hat{J}_2$,
\begin{equation}
\label{eq2:ddphiJ2}
\frac{\partial}{\partial \phi} \langle \hat{J}_2 \rangle =
\langle \hat{J}_1 \rangle =
-i(\langle \psi \mid \! \hat{J}_2 ~\hat{J}_3  \! \mid  \psi \rangle -
\langle \psi \mid \! \hat{J}_3 ~\hat{J}_2 \! \mid  \psi \rangle).
\end{equation}
That is, the phase derivative of $\langle \hat{J}_2 \rangle$ is
equal to twice the imaginary part of the inner product of
$\hat{J}_2\mid \psi \rangle$ and $\hat{J}_3\mid \psi \rangle$.
For $\langle \hat{J}_2 \rangle =0 $ and
$\langle \hat{J}_3 \rangle =0 $, the lengths of these two
Hilbert space vectors are given by the uncertainties $\Delta J_2$
and $\Delta J_3$. Since the inner product of two vectors cannot be
larger than the product of the lengths of the two vectors, the
uncertainty product is limited by the Cauchy-Schwartz inequality
\begin{equation}
\label{eq2:uncertaintyJ2J3}
\Delta J_2 \Delta J_3 \geq \frac{1}{2}
\frac{\partial}{\partial \phi} \langle \hat{J}_2 \rangle.
\end{equation}
This is just a specific case of the Mandelstam-Tamm uncertainty relations that limit
the phase sensitivities of quantum states \cite{Bra96}.
A quantum state that achieves the limit of this uncertainty
relation achieves its optimal phase sensitivity - the
quantum Cramer-Rao bound \cite{Gio06,Bra94} -
with $\hat{J}_2$ as its optimal phase estimator.
By substituting eq.(\ref{eq2:deltaphi}) into
eq.(\ref{eq2:uncertaintyJ2J3}), it can be
confirmed that the phase sensitivity limit is indeed
equal to the well known bound given by the
uncertainty of the generator $\hat{J}_3$,
\begin{equation}
\label{eq2:minideltaphi}
\delta \phi^2 \geq \frac{1}{4 \Delta J_3^2}.
\end{equation}
The implications of this relation for optical quantum metrology have
been explained with great clarity in \cite{Gio06}. In particular,
eq.(\ref{eq2:minideltaphi}) shows that the maximal phase
sensitivity is obtained for path entangled states, which have
a maximal possible $\hat{J}_3$-uncertainty of $\Delta J_3^2
= N^2/4$ and can therefore achieve the Heisenberg limit (HL) of
$\delta \phi^2 = 1/N^2$. On the other hand, uncorrelated photons
have a $\hat{J}_3$-uncertainty of $\Delta J_3^2 = N/4$,
corresponding to a completely random distribution between the
two paths of the interferometer. Non-entangled photons can therefore
only achieve the standard quantum limit (SQL) of
$\delta \phi^2 = 1/N$.

\begin{figure}[t]
\begin{center}
\includegraphics[scale=1.0]{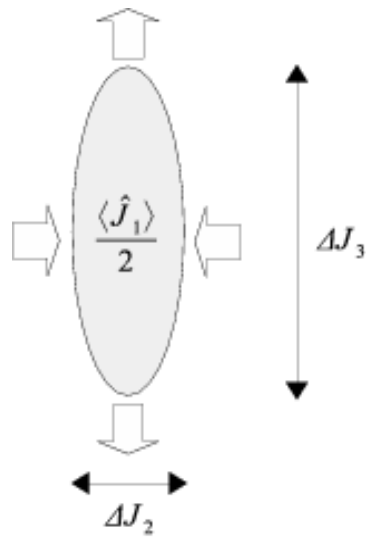}
\caption{\label{squeeze}
Illustration of squeezing in the $J_2$-$J_3$ plane. A reduction
of the output uncertainty $\Delta J_2$ requires an increase in
path uncertainty $\Delta J_3$, so that the uncertainty
product remains equal to $\langle \hat{J}_1 \rangle/2$.
}
\end{center}
\end{figure}

Having established the consistency of our analysis with more general
quantum metrology, we can now return to our specific phase
estimation strategy characterized by the uncertainty vectors
$\hat{J}_2\mid \psi \rangle$ and $\hat{J}_3\mid \psi \rangle$.
According to eq.(\ref{eq2:ddphiJ2}), the phase sensitivity of
the estimator observable $\hat{J}_2$ is optimal if the
vectors $\hat{J}_2\mid \psi \rangle$ and $\hat{J}_3\mid
\psi \rangle$ have the same direction in Hilbert space with
a purely imaginary inner product. If the ratio of the
uncertainties in $\hat{J}_2$ and $\hat{J}_3$ is given by a
squeezing factor of $\exp(-2 r)$, this condition can be expressed
by
\begin{equation}
\label{eq2:optimalcondition}
\mathrm{e}^{r}\hat{J}_2 \mid \psi \rangle =
i\mathrm{e}^{-r} \hat{J}_3 \mid \psi \rangle.
\end{equation}
Eq.(\ref{eq2:optimalcondition}) defines the complete class of
$N$-photon squeezed states for which $\hat{J}_2$ is the optimal phase
estimator. Specifically, these states are all minimal uncertainty states
of the $\hat{\bf J}$-vector components $\hat{J}_2$ and $\hat{J}_3$, where
the $\hat{J}_2$-uncertainty is squeezed and the
$\hat{J}_3$-uncertainty
is anti-squeezed, as illustrated in fig. \ref{squeeze} \cite{Hil93}.
Since the uncertainty product $\Delta J_2 \Delta J_3$ for these
states is equal to its minimal value of $\langle \hat{J}_1 \rangle/2$,
the uncertainties can be given separately as
\begin{equation}
\label{eq2:deltaJ2J3}
\Delta J_2^2 = \frac{\mathrm{e}^{-2r}}{2} \langle \hat{J}_1 \rangle,
\hspace{1cm} \Delta J_3^2 =
\frac{\mathrm{e}^{2r}}{2} \langle \hat{J}_1 \rangle.
\end{equation}
According to eq.(\ref{eq2:deltaphi}), the phase sensitivity of
these states is then defined by the minimal phase error allowed by
the uncertainty relation given by eq.(\ref{eq2:minideltaphi}),
\begin{equation}
\label{eq2:sqdeltaphi}
\delta \phi^2 = \frac{\mathrm{e}^{-2r}}{\langle \hat{J}_1 \rangle}.
\end{equation}
Since this phase error cannot be lower than the HL of $1/N^2$, we can
expect that $\langle J_1 \rangle$ will drop to zero with $\exp(-2r)$.
On the other hand, at squeezing levels well below the HL, we
can expect $\langle J_1 \rangle$ to be very close to $N/2$, so that
squeezing factors of $\exp(-2 r)\gg 1/N$ directly describe the suppression of the phase error.

In the completely unsqueezed limit of $r=0$,
eq.(\ref{eq2:optimalcondition}) actually defines the
$\hat{J}_1$ eigenstate $\mid N;0 \rangle$ with the maximal
eigenvalue of $N/2$, where all photons enter the Mach-Zehnder
interferometer in mode $\hat{a}$. Experimentally,
this state is easily realized by using coherent laser light to
generate the single mode input in $\hat{a}$, while mode $\hat{b}$
is left in the vacuum state.
The uncertainty vectors of this ``classical''
limit of quantum metrology are given by
\begin{equation}
\label{eq2:J2J3}
\hat{J}_2 \mid N;0 \rangle = \sqrt{\frac{N}{4}} \mid N-1;1 \rangle
= i\hat{J}_3 \mid N;0 \rangle.
\end{equation}
Therefore, the uncertainties in $\hat{J}_2$ and $\hat{J}_3$ are
both equal to $\sqrt{N/4}$, which corresponds to the shot noise
expected from a random distribution of the $N$ input photons
between the paths and the output ports. The phase error of this
``classical'' input light is then given by the SQL of
$\delta \phi^2 = 1/N$, defining the phase sensitivity limit
of completely uncorrelated photons.

In the following, we will look into the possibility of generating
$N$-photon squeezed light that beats the SQL by quantum interference
of laser light with only a small fraction of down-converted light.
The resulting phase sensitivity will then be somewhere
between the SQL and the HL. To evaluate just how strong the
non-classical effects are, it is convenient to define the quantum
enhancement parameter $Q$ with
\begin{equation}
\label{eq2:Q}
Q = \frac{\ln (1/(N \delta \phi^2))}{\ln (N)}.
\end{equation}
This parameter describes the logarithmic phase sensitivity
on a scale from $0$ for the SQL to $1$ for the HL and thus
provides a convenient tool for comparing the quantum enhancements
achieved at different photon numbers $N$.

\section{Generation of $N$-photon squeezed states
by quantum interference}
\label{sec:qi}
As explained in the previous section, the phase estimation error
$\delta \phi^2$ originates from the uncertainty in the output photon
number difference $\hat{J}_2$. In Hilbert space, this uncertainty
is described by a vector $\hat{J}_2 \mid \psi \rangle$.
It is therefore possible to reduce the uncertainty in $\hat{J}_2$
by adding a quantum state component whose uncertainty vector points
in the opposite direction in Hilbert space. The uncertainties of
the two components then interfer destructively, resulting in
an $N$-photon squeezed state with a $\hat{J}_2$-uncertainty that is lower
than that of either one of the components.

For the $\mid \! N;0 \rangle$ state describing the laser input
in mode $\hat{a}$, the uncertainty vector is given by
\begin{equation}
\label{eq:SQLstate}
\hat{J}_2 \mid \! N;0 \rangle = \sqrt{\frac{N}{4}}
\mid \! N-1;1 \rangle.
\end{equation}
To reduce the uncertainty by quantum interference, we need
an orthogonal state with an uncertainty component of
$\mid \! N-1;1 \rangle$. Since the application of
$\hat{J}_2$ exchanges exactly one photon between the
input modes $\hat{a}$ and $\hat{b}$, the
$\mid \! N-2;2 \rangle$ state satisfies this requirement.
Specifically,
\begin{equation}
\label{eq:QIstate}
\hat{J}_2 \mid \! N-2;2 \rangle = \sqrt{\frac{N-1}{2}}
\mid \! N-1;1 \rangle + \sqrt{\frac{3(N-2)}{4}}
\mid \! N-3;3 \rangle.
\end{equation}
The $\mid \! N-1;1 \rangle$-component of this uncertainty
vector has the same sign as the uncertainty in eq.
(\ref{eq:SQLstate}), so a negative superposition of the
shot noise limited laser light input $\mid \!N;0 \rangle$
and the state $\mid \!N-2;2 \rangle$ with two photons in mode
$\hat{b}$ will result in a reduction of the $\hat{J}_2$
uncertainty to below the SQL.

As has been shown in a number of papers \cite{Sha04,Liu06,Lu02}, quantum
interference effects can be obtained by combining weak
coherent laser
light with down-converted photon pairs. Just as in other
down-conversion based experiments, the $N$-photon component
is then selected by selectively detecting the $N$-photon
coincidences in the output ports.
In the present case,
a quantum coherent superposition of the $\mid \!N;0 \rangle$
state and the $\mid \!N-2;2 \rangle$ can be realized by
adding weak parametric down-conversion (PDC) in mode $\hat{b}$
to the coherent laser light in mode $\hat{a}$.
If the emission probability of photon pairs from PDC is very small, the down-converted light is approximately represented by the
addition of a very small photon pair component to the vacuum state,
\begin{equation}
\label{eq3:PDC}
\mid \gamma \rangle \approx \mid 0 \rangle -
\frac{\gamma}{\sqrt{2}} \mid 2 \rangle,
\end{equation}
where $\gamma \ll 1$. The remaining $N$ or $N-2$ photons originate
from the coherent laser light in mode $\hat{a}$. The relevant
components of the coherent state are given by
\begin{equation}
\label{eq3:coherent}
\mid \alpha \rangle = \mathrm{e}^{-\frac{|\alpha|^2}{2}}(\ldots +
\frac{\alpha^{N-2}}{\sqrt{(N-2)!}} \mid N-2 \rangle \ldots
+ \frac{\alpha^{N}}{\sqrt{N!}} \mid N \rangle \ldots).
\end{equation}
We can then obtain the $N$-photon component of the product state
of eq.(\ref{eq3:PDC}) and eq.(\ref{eq3:coherent}),
\begin{equation}
\label{eq3:N0N2state}
\mid \psi \rangle \approx \mid N;0 \rangle - \sqrt{\frac{N(N-1)}{2}}\frac{\gamma}{\alpha^2} \mid N-2;2 \rangle,
\end{equation}
where the probability amplitude of $\sqrt{N(N-1)/2}(\gamma/\alpha^2)
\approx (N \gamma/\alpha^2)/\sqrt{2}$ should be much smaller than
one. Note that the phase of the superpostion is controlled by the
phase relation between the down-conversion amplitude $\gamma$ and
the squared coherent state amplitude $\alpha^2$. Quantum interference
between the two light sources is possible because the origin of the
photons cannot be distinguished in the output measurements.
The effect of this quantum interference on the uncertainty of the output
photon number difference $\hat{J}_2$ is given by
\begin{equation}
\label{eq3:1N0N2deltaJ2}
\Delta J_2^2 \approx \langle N;0 \mid \hat{J}_2^2
\mid N;0 \rangle - 2 \sqrt{\frac{N(N-1)}{2}}\frac{\gamma}{\alpha^2}
\langle N-2;2 \mid \hat{J}_2^2  \mid N;0 \rangle.
\end{equation}
The precise amount of noise reduction in the output is determined
by the quantum interference term $\langle N-2;2 \mid
\hat{J}_2^2  \mid N;0 \rangle$ of the operator $\hat{J}_2^2$.
Using the appropriate matrix elements of $\hat{J}_2$, the result is
\begin{equation}
\label{eq3:2N0N2deltaJ2}
\Delta J_2^2 \approx \frac{N}{4}(1 - 2 (N-1)\frac{\gamma}{\alpha^2}).
\end{equation}
Since the change in $\langle \hat{J}_1 \rangle$ is only of the
order of $\gamma^2$, this reduction in the uncertainty of
$\hat{J}_2$ translates directly into a corresponding reduction of
the phase error. The quantum enhancement parameter $Q$ is then
given by
\begin{equation}
\label{eq3:Qweak}
Q \approx 2 \frac{(N-1)}{\ln(N)} \frac{\gamma}{\alpha^2}.
\end{equation}
Significantly, the quantum enhancement increases with the
total photon number $N$, indicating that the selection of
the $N$-photon component in the output plays an important
role in the achievement of high squeezing levels.

\section{Characterization of highly squeezed
$N$-photon states}
\label{sec:operators}
In the previous section, we explained the basic effect of squeezing
by quantum interference in the limit of very weak down-conversion,
where the generation of multiple photon pairs can be neglected.
However, the effect of multiple photon pairs in the down-converted
light may be important for the generation of highly squeezed light.
In this section, we therefore derive a more general expression
for the interferometric squeezing effect based on the relation
of the uncertainty vectors of $\hat{J}_2$ and $\hat{J}_3$ given
by eq.(\ref{eq2:optimalcondition}) in section \ref{sec:MZ}.

In general, the quantum state $\mid \! \gamma \rangle$ describing
down-converted light is a slightly squeezed vacuum state.
It is therefore possible to describe it as right eigenstate of
a squeezed annihilation operator
$(\hat{b} + \gamma \hat{b}^{\dagger})$
with an eigenvalue of zero.
Note that the parameter $\gamma$
describing the quadrature squeezing is usually very close to zero,
since the quadrature squeezing achieved in conventional
down-conversion is negligibly small.
In the present context,
the quadrature squeezing relation is used to characterize the
quantum coherence of the down-converted light, which is described
by the operator relation
\begin{equation}
\label{eq4:bPDC}
\hat{b} \mid \! \gamma \rangle = -\gamma \hat{b}^{\dagger} \mid \! \gamma \rangle.
\end{equation}
Likewise, the coherent state $\mid \! \alpha \rangle$ can be defined
as the right eigenstate of its annihilation operator $\hat{a}$
with eigenvalue $\alpha$,
\begin{equation}
\label{eq4:aalpha}
\hat{a} \mid \! \alpha \rangle = \alpha \mid \! \alpha \rangle.
\end{equation}
We can now combine these two relations to characterize the coherence
of the product state input $\mid \! \alpha;\gamma \rangle$.
However, this state is still a superposition of various
total photon numbers $N$. Since we are interested in
the results of coincidence counting experiments that select only the $N$-photon component of this state, where $N$ is usually much higher than the average photon number, it is
necessary to separate the coherence of the $N$-photon
component from the (generally different) quantum coherences
at other photon numbers. We have found that this problem
can be solved by formulating an operator relation using
only operators that do not
change the total photon number, such as $\hat{a}^{\dagger}\hat{b}$
and $\hat{a}\hat{b}^{\dagger}$.
The most simple relation we can
thus derive is
\begin{eqnarray}
\label{eq4:abeta}
\hat{a}^{\dagger}\hat{b} \mid \! \eta \rangle &=&
-\frac{\gamma}{\alpha^2}
\hat{a}^{\dagger}\hat{a}\hat{a}\hat{b}^{\dagger} \mid \! \eta
\rangle\nonumber\\ &=& -\eta
\frac{(N-\hat{b}^{\dagger}\hat{b})}{N}\hat{a}\hat{b}^{\dagger}
\mid \! \eta \rangle,
\end{eqnarray}
where $\mid \eta \rangle$ is the $N$-photon component detected
in a corresponding coincidence counting experiment.
The single parameter $\eta$ that defines the properties of
this $N$-photon squeezed state is given by
\begin{equation}
\label{eq4:eta}
\eta = \frac{N \gamma}{\alpha^2}.
\end{equation}
Here, the definition of $\eta$ has been chosen so that the
form of eq.(\ref{eq4:abeta}) defining the $N$-photon
state $\mid \eta \rangle$ is similar to the definition
of the squeezed vacuum in eq.(\ref{eq4:bPDC}). As a result,
the $N$-dependence of $\eta$ properly expresses the fact that
the squeezing levels of the $N$-photon components in the
same input state $\mid \! \alpha;\gamma \rangle$ increase
with $N$.

Using the Schwinger representation, we can now express the
coherence operators $\hat{a}^{\dagger}\hat{b}$
and $\hat{a}\hat{b}^{\dagger}$ in terms of the output operator
$\hat{J}_2$ and the path operator $\hat{J}_3$. We then obtain
a non-linear version of the squeezing relation given
by eq.(\ref{eq2:optimalcondition}),
\begin{equation}
\label{eq4:generaleta}
(1 + \frac{(N - \hat{b}^{\dagger}\hat{b})}{N} \eta)\hat{J}_2 \mid \!
\eta \rangle = i(1 - \frac{(N -
\hat{b}^{\dagger}\hat{b})}{N} \eta) \hat{J}_3 \mid \! \eta \rangle.
\end{equation}
Eq.(\ref{eq4:generaleta}) is an exact definition of the $N$-photon
states generated by the interference of coherent laser light and
down-converted light. However, the non-linear dependence of the
quantum state on the photon number $\hat{b}^\dagger \hat{b}$
in the down-converted input beam introduces features that are
quite different from the squeezing effects described
in section \ref{sec:MZ} as $\eta$ increases \cite{Hof07}.
In the following, we will therefore focus on the case of
small $\eta$, where most of the photons originate from the
coherent laser light input. We can then assume that
$\langle \hat{b}^{\dagger}\hat{b} \rangle \ll N$, so that
the approximate equation for the $N$-photon squeezed state
$\mid \! \eta \rangle$ can be written as
\begin{equation}
\label{eq4:sqeta}
(1 + \eta)\hat{J}_2 \mid \! \eta \rangle
\approx i(1 - \eta)\hat{J}_3 \mid \! \eta \rangle.
\end{equation}
Except for a constant factor, this equation is equal to
eq.(\ref{eq2:optimalcondition}), and therefore describes the
minimal uncertainty state introduced in section \ref{sec:MZ}
with a squeezing factor of
\begin{equation}
\label{eq4:reta}
\mathrm{e}^{-2 r} = \frac{\Delta J_2}{\Delta J_3} = \frac{1-\eta}{1+\eta}.
\end{equation}
Thus, the $N$-dependent parameter
$\eta=N \gamma/\alpha^2$ directly defines the
level of squeezing for the $N$-photon component
in the limit where most of the $N$ photons originate
from the coherent light input.
In particular, the $N$-dependence of the squeezing
parameter $\eta$
means that the squeezing properties of the input
light are amplified by selecting a component with
a photon number $N$ that is
much greater than the average photon number of the input
state, as it is presently done in typical down-conversion
based experiments \cite{Wal04,Mit04,Eis05,Lu02}.
Our result thus shows how the selection of $N$ photons
in coincidence counting can result in strong non-classical
effects, even though the actual squeezing level of
down-converted light is negligible ($\gamma \ll 1$).

\section{Limits of interferometric squeezing}
\label{sec:approx}

Eq.(\ref{eq4:reta}) suggests that arbitrarily high
squeezing levels can be obtained as $\eta$ approaches
one. However, it is clear that the approximation that
most photons in the $N$-photon state $\mid \eta \rangle$ originate from the coherent light input breaks down
well before $\eta$ reaches one. It is therefore interesting
to consider the limits of the approximation in order
to determine the maximal squeezing levels that can be
obtained by quantum interference between coherent light
and down-converted light.

Since the approximation used to derive eq.(\ref{eq4:sqeta})
is based on the assumption that the number of
down-converted photons in the $N$-photon state
$\mid \eta \rangle$ is negligibly small, the limit
of this approximation can be found by determining
the approximate relation
between the squeezing
parameter $\eta$ and the average photon number in mode $\hat{b}$.
To do so, we consider the
square of the $\hat{\bf J}$-vector, which is given by
$\hat{J}_1^2 + \hat{J}_2^2 + \hat{J}_3^2 = (N/2)(N/2 + 1)$.
Since $\hat{J}_1=N/2-\hat{b}^\dagger \hat{b}$, we can use
the approximation $\langle \hat{J}_1^2 \rangle \approx (N/2)^2
- N \langle \hat{b}^{\dagger} \hat{b} \rangle$ to derive the
relation
\begin{equation}
\label{eq4:btbest1}
\langle \hat{J}_2^2 \rangle + \langle \hat{J}_3^2 \rangle \approx N\left(\langle \hat{b}^{\dagger}\hat{b} \rangle + \frac{1}{2}
\right).
\end{equation}
The expectation values $\langle \hat{J}_2^2 \rangle$ and
$\langle \hat{J}_3^2 \rangle$ are equal to the
uncertainties given by eq.(\ref{eq2:deltaJ2J3}),
with $\langle \hat{J}_1 \rangle \approx N/2$ and
$\exp(-2r)=(1-\eta)/(1+\eta)$. The approximate relation
between the squeezing parameter $\eta$ and the average photon
number in input mode $\hat{b}$ thus reads
\begin{equation}
\label{eq4:btbest2}
\langle \hat{b}^{\dagger}\hat{b} \rangle \approx
\frac{\eta^2}{1-\eta^2}.
\end{equation}
For $\eta\ll 1$, the average photon number in mode $\hat{b}$
increases with $\eta^2$, as expected from the approximate result
in eq.(\ref{eq3:N0N2state}) of section \ref{sec:qi}.
Large photon numbers are only obtained
when $\eta$ is close to one. We can therefore estimate that
eq.(\ref{eq4:sqeta}) is valid until $(1-\eta)$ is considerably
smaller than one.

It is now possible to get a more precise idea of where the
deviations between the approximate definition of
$\mid \eta \rangle$ in eq.(\ref{eq4:sqeta}) and the exact
definition in eq.(\ref{eq4:generaleta}) become relevant.
Specifically, the factor of
$(1-\eta+\eta \hat{b}^\dagger \hat{b}/N)$ in the right hand side of
eq.(\ref{eq4:generaleta}) is approximated by $(1-\eta)$ in
the right hand side of eq.(\ref{eq4:sqeta}).
Thus, the assumption is that
\begin{equation}
\label{eq4:etaest1}
1-\eta > \frac{\eta}{N} \langle \hat{b}^\dagger \hat{b} \rangle.
\end{equation}
For $(1-\eta)\ll 1$, the factor of $\eta$ on the right hand
side of this relation is approximately equal to one.
The same approximation can also be used to simplify the
expression for the average photon number
from down-conversion given
by eq. (\ref{eq4:btbest2})
to $\langle \hat{b}^\dagger \hat{b} \rangle \approx
1/(2(1-\eta))$. It is then possible to give the
condition for the approximation used in eq.(\ref{eq4:sqeta})
as
\begin{equation}
\label{eq4:etaest2}
1-\eta > \frac{1}{\sqrt{2 N}}.
\end{equation}
We can use this condition to derive estimates of the
maximal average number of
down-converted photons and of the minimal
phase error achieved at the point where the
approximation breaks down from
eqs.(\ref{eq4:btbest2}) and (\ref{eq4:reta}).
The results read
\begin{eqnarray}
\label{eq4:btbmax}
\langle \hat{b}^\dagger \hat{b} \rangle &<& \sqrt{\frac{N}{2}},
\\
\label{eq4:deltaphimin}
\delta \phi^2 &>& \frac{1}{(2 N)^{3/2}}.
\end{eqnarray}
Eq.(\ref{eq4:btbmax}) indicates that the
optimal squeezing levels will be achieved when
the average photon number in mode $\hat{b}$ is
close to $\sqrt{N/2}$.
At high photon numbers $N$, this is still only a small
fraction of the total photon number. We can therefore
conclude that most photons originate from the
coherent laser light, even when the squeezing is
maximal.

Eq.(\ref{eq4:deltaphimin}) provides an estimate of the amount of
phase squeezing that can be achieved. Since it has been obtained
by extrapolating the approximation of eq.(\ref{eq4:sqeta}) to the
point where it breaks down, we can assume that the actual minimum
of the phase error is a little bit higher than the lower bound
given by eq.(\ref{eq4:sqeta}).
In fact, numerical simulations such as the one
presented in section \ref{sec:numerics} indicate that
it is reasonable to expect the actual minimum phase
error around $\delta\phi^2 = 1/N^{3/2}$, which is equal
to the geometric mean
of the HL and the SQL.
In terms of the quantum enhancement factor, the
limit of phase sensitivity achieved by interferometric
squeezing is given by
\begin{equation}
\label{eq4:Qhalf}
Q_{\mbox{max.}} \leq \frac{1}{2}.
\end{equation}
The interference of laser light and down-converted photon
pairs thus produces $N$-photon squeezed states with quantum
enhancement factors of $Q=0$ to $Q=1/2$ for values of $\eta=0$
to $\eta=1$.

It may be worth noting that this result corresponds to
the maximal phase sensitivity obtained by using interference
between a squeezed vacuum and coherent light without
selecting a specific $N$-photon component \cite{Par95}.
In that case, $N$ gives only the average photon number
and the squeezing level is given directly by $\gamma$
instead of $\eta$. Thus, it is not immediately obvious
that this result can also be applied to the actual $N$-photon
component $\mid \eta \rangle$. However, it is possible to
understand the similarity with our result by noting
that, for average values of $N$ much greater than one,
the photon numbers with the highest
probabilities occur close to $N \approx \alpha^2$, where
the $N$-photon squeezing parameter is $\eta \approx \gamma$.
Thus the average phase sensitivity over all $N$ is close
to the phase sensitivity observed for the special case
of $N \approx \alpha^2$.
By focussing on the quantum statistics of a specific
$N$-photon component, our theory shows that the squeezing
limit of $\delta\phi^2 = 1/N^{3/2}$ applies not only
to a high average photon number of $N \gg 1$, but
also to the high $N$ components of extremely weak
non-classical light observed by coincidence counting,
where the squeezing parameter is
increased by a factor of $N/\alpha^2 \gg 1$.

Since our theory provides a more detailed description
of the $N$-photon component, it also allows us to identify
the kind of $N$-photon states generated at $\eta>1$.
As we have shown elsewhere \cite{Hof07}, the state generated
at $\eta=2$ is actually close to a path
entangled state, with a fidelity of 94 \% at high photon numbers $N$.
We can therefore conjecture that eq.(\ref{eq4:generaleta})
describes a transition from squeezing to a quantum
superposition of separate regions on the $\hat{\bf J}$-vector
sphere at $\eta=1$.
As was recently pointed out by Pezze and Smerzi \cite{Pez07},
it is possible to achieve the quantum Cramer-Rao bound
for such states (and thus the Heisenberg limit at $\eta=2$)
by optimizing the phase estimation method used to evaluate the
output of the two path interferometer.
This means that the phase sensitivity of the states actually
continues to increase for $\eta>1$. However, the linear phase
estimation procedure characterized by eq.(\ref{eq2:deltaphi})
will not be optimal anymore, so that the high phase sensitivity
can only be observed by using a more elaborate and
error-sensitive phase estimation procedure.

\section{Generation probability of the $N$-photon $\eta$-state}
\label{sec:state}

The two main advantages of the $N$-photon squeezed
state $\mid \eta \rangle$ obtained by interference of laser light and down-converted
photon pairs are that it permits a fairly straightforward and
error-resistant phase estimation, and that it can be generated
at high efficiency since most of the photons originate from
the laser light input. In order to analyze the latter point, it
is necessary to consider the complete quantum statistics of the
interference between laser light and down-converted photon pairs.
Since the photon number expansions of both the coherent state
$\mid \alpha \rangle$ and the squeezed vacuum $\mid \gamma \rangle$
are well known, it is a straightforward matter to write out the
complete product state in the input basis. The result is a
superposition of outputs with different photon numbers $N$ given by
\begin{equation}
\hspace{-2.5cm}
\label{eq5:alphagamma}
\mid \alpha; \gamma \rangle =
(1-\gamma^2)^{1/4} \mathrm{e}^{-|\alpha|^2/2}
\sum_{N=0}^{\infty} \frac{\alpha^N}{\sqrt{N!}}
\sum_{k=0}^{N/2} \frac{1}{k!}\sqrt{\frac{(2k)! N!}{(N-2k)!}}
\left(\frac{\gamma}{2 \alpha^2} \right)^k \mid N-2k, 2k \rangle.
\end{equation}
The total probability of obtaining an $N$-photon squeezed state
is then given by
\begin{eqnarray}
\label{eq5:Psq}
P_{\mbox{sq.}}(N) &=&
\sqrt{1-\gamma^2} \mathrm{e}^{-|\alpha|^2}
\frac{|\alpha|^{2N}}{N!} \frac{1}{|C_N|^2},
\nonumber \\
\mbox{where} &&
\frac{1}{|C_N|^2} = \sum_{k=0}^{N/2}
\frac{(2k)! N!}{(k!)^2 (N-2k)!}
\left| \frac{\gamma}{2 \alpha^2} \right|^{2k}.
\end{eqnarray}
The constant $C_N$ can also be used to express the normalized
$N$-photon state $\mid \! \eta \rangle_N$, which is the state
defined by eq.(\ref{eq4:generaleta}) with $\eta=N \gamma /\alpha^2$,
\begin{equation}
\label{eq5:etastate}
\mid \eta \rangle_N = C_N
\sum_{k=0}^{N/2} \frac{1}{k!}\sqrt{\frac{(2k)! N!}{(N-2k)!}}
\left(\frac{\eta}{2 N} \right)^k \mid N-2k; 2k \rangle.
\end{equation}
Here, $C_N$ is defined as the probability amplitude of the
$\mid N;0 \rangle$ component of the $N$-photon $\eta$-state.
It may be worth noting that $C_N$ depends only on the value of the squeezing parameter $\eta$ and the total photon number $N$, so it can
be determined without knowing the precise values of $\alpha$
and $\gamma$.

In previous experiments investigating the enhanced phase
sensitivity of $N$-photon states, down-conversion was the
only photon source used \cite{Mit04,Eis05,Sun06,Nag07,Sun07}.
As a result, the $N$-photon
coincidence count rate was limited by the available
down-conversion amplitude $\gamma$, which is usually much
smaller than one.
In the case of interferometric squeezing discussed in this
paper, the source of the non-classicality is also a
down-conversion process of amplitude $\gamma$. However,
most of the $N$-photons of the squeezed states originate
from the coherent laser light. It is therefore possible
to achieve high $N$-photon coincidence rates, even if the down-converted amplitude $\gamma$ used is rather small.
Since the technological requirements of interferometric
squeezing are otherwise very similar to those for
$N$-photon down-conversion, it is interesting to compare the
$N$-photon squeezed state generation probability with
the corresponding generation probability of an $N$-photon
pair state $\mid N/2;N/2\rangle$ such as the one used in
the recent demonstrations of phase super-resolution
\cite{Nag07,Sun07}.

When the down-conversion amplitude $\gamma$ is much smaller than one, the approximate generation probability for a pair state is given by
\begin{equation}
\label{eq5:Ppair}
P_{\mbox{\small pair}}(N) = (1-\gamma^2)\gamma^N \approx \gamma^N.
\end{equation}
Thus, the multi-photon generation probability
falls off exponentially with $N$, indicating that the low down-conversion amplitude severely limits the efficiency of
multi-photon pair state generation.
The corresponding $N$-photon generation probability of an
interferometric squeezed state is given by
eq.(\ref{eq5:Psq}). For a fixed
down-conversion amplitude $\gamma$, it is possible to achieve
any value of the squeezing parameter $\eta$, simply by varying
$\alpha$ according to eq.(\ref{eq4:eta}).
Specifically, the coherent amplitude needed to obtain a
given value of the squeezing parameter $\eta$ is
$\alpha=\sqrt{N \gamma/\eta}$.
Using the assumption that both $\alpha$ and $\gamma$ are
much smaller than one, the approximate $N$-photon generation probability then reads
\begin{equation}
\label{eq5:Psqapprox}
P_{\mbox{\small sq.}}(N) \approx
\frac{1}{N!}(\frac{N\gamma}{\eta})^N \frac{1}{|C_N|^2}
\approx \frac{1}{\sqrt{2\pi N}} (\frac{\mathrm{e} \gamma}{\eta})^N
 \frac{1}{|C_N|^2},
\end{equation}
where the second approximation is made using the Sterling formula
$N! \approx \sqrt{2\pi N}\mathrm{e}^{-N}N^N$ for large $N$.
Eq.(\ref{eq5:Psqapprox}) shows that the $\eta$-state generation
probability also falls off exponentially
with photon number $N$. However, the exponential decline of the $\eta$-state coincidence count rate
is slower than $N$-photon down-conversion by a factor of $\mathrm{e}/\eta$ in the base. The ratio of
the $N$ photon generation probabilities of the squeezed
state in eq.(\ref{eq5:Psq}) and of the pair state in eq.(\ref{eq5:Ppair}) is therefore independent of $\gamma$
and reads
\begin{equation}
\label{eq5:PsqPpair}
\frac{P_{\mbox{\small sq.}}(N)}{P_{\mbox{\small pair}}(N)} \approx \frac{1}{\sqrt{2\pi N}}
\left(\frac{\mathrm{e}}{\eta} \right)^N \frac{1}{|C_N|^2}.
\end{equation}
Thus, the relative increase of the coincidence count rate
compared to pair states is approximately exponential with photon number, indicating that interferometric squeezing is an especially promising method for increasing the number of photons N observed in coincidence counting.

\begin{table}
\caption{Comparison of $N$-photon squeezed state generation probabilities $P_{\mathrm{sq.}}$ with the generation probabilities $P_{\mathrm{pair}}$ for $N$-photon pair states using the same down-conversion source for photon number from N=3 to N=8. The three columns of the table show the improvement of coincidence count rates expected for $\eta=1/3$, $\eta=1/2$ and $\eta=1$ respectively. The numbers in parenthesis show the suppression of phase error below the standard quantum limit.}
\label{countrate}
\begin{center}
\begin{tabular}{c|cc|cc|cc|}
&\multicolumn{2}{|c|}{$\eta=1/3$}&\multicolumn{2}{|c|}{$\eta=1/2$}&
\multicolumn{2}{|c|}{$\eta=1$}\\
$N$&$P_{\mathrm{sq.}}/P_{\mathrm{pair}}$&($N \delta \phi^2$)&$P_{\mathrm{sq.}}/P_{\mathrm{pair}}$&($N \delta \phi^2$)&$P_{\mathrm{sq.}}/P_{\mathrm{pair}}$&($N \delta \phi^2$)\\
\hline
& & & & & &\\[-0.2cm]
3&130&(0.68)&40&(0.60)&6&(0.75)\\
4&$0.9 \times 10^3$&(0.64)&$1.9 \times 10^2$&(0.55)&$15$&(0.57)\\
5&$0.7 \times 10^4$&(0.62)&$0.9 \times 10^3$&(0.51)&$39$&(0.51)\\
6&$0.5 \times 10^5$&(0.60)&$0.5 \times 10^4$&(0.49)&$100$&(0.45)\\
7&$3.8 \times 10^5$&(0.58)&$2.4 \times 10^4$&(0.47)&$260$&(0.41)\\
8&$2.9 \times 10^6$&(0.57)&$1.2 \times 10^5$&(0.45)&$680$&(0.37)\\
\end{tabular}
\end{center}
\end{table}


To illustrate the improvement in the coincidence count rate over pair state generation, table \ref{countrate} shows the ratio $P_{\mathrm{sq.}}/P_{\mathrm{pair}}$ for the experimentally relevant range
of photon numbers from 3 to 8 photons. The ratios have been determined
from eq.(\ref{eq5:Psq}) and eq.(\ref{eq5:Ppair}) without any approximations.
Of particular interest might be the case of $N=4$, since four photon
coincidences have been used in the recent experiments demonstrating
phase super-sensitivity \cite{Nag07,Sun07}. In this case, beating the
SQL with a highly squeezed four photon state at $\eta=1$ would
already improve the coincidence count rate by a factor of 15. By reducing
the squeezing parameter $\eta$ to $1/3$, the count rate
can even be increased to 900 times that of pair state generation, at a
phase error that is still lower than the SQL by a factor of $0.64$.
In general, the results in table \ref{countrate} show that the coincidence
count rates of interferometric squeezed states should be several
orders of magnitude higher than those obtained in
pair state generation, even at presently accessible photon numbers.


\section{Numerical results for the eight photon squeezed state}
\label{sec:numerics}
To confirm the validity of the approximations used in the previous sections and to
take a closer look at the maximally squeezed state, we now investigate the specific
numerical results that can be obtained for the $N=8$ case.
According to eq.(\ref{eq5:etastate}), the eight photon state is given by
\begin{eqnarray}
\label{eq6:8state}
\fl
\mid \! \eta \rangle_{N=8} =
 C_8 \left( \mid \! 8;0 \rangle - \frac{\sqrt{7}}{4} \eta
\mid \! 6;2 \rangle + \frac{3 \sqrt{70}}{64} \eta^2 \mid \! 4;4 \rangle
- \frac{15 \sqrt{7}}{256} \eta^3 \mid \! 2;6 \rangle + \frac{105}{4096} \eta^4
\mid \! 0;8 \rangle \right)
\nonumber \\
\fl \hspace*{1.4cm}
\approx  C_8 \left( \mid \! 8;0 \rangle - 0.661 \eta
\mid \! 6;2 \rangle + 0.392 \eta^2 \mid \! 4;4 \rangle
- 0.155 \eta^3 \mid \! 2;6 \rangle + 0.0256 \eta^4
\mid \! 0;8 \rangle \right).\nonumber\\
&&
\end{eqnarray}
In this expansion, the first two terms correspond exactly to the $N=8$ case of the
approximation given in eq.(\ref{eq3:N0N2state}). Below $\eta=1/2$, these two terms
alone contribute more than 99 \% of the state. Above $\eta=1/2$, higher order terms
become relevant and the approximations for low $\eta$ cease to apply.

\begin{figure}[t]
\begin{center}
\includegraphics[scale=1.0]{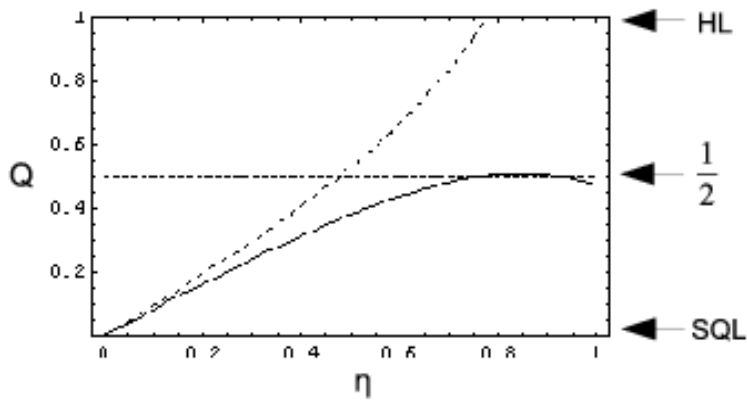}
\vspace{-0.3cm}
\caption{\label{Qparameter}
Quantum enhancement parameter $Q$ describing the phase
sensitivity achieved by the
eight photon state as a function of $\eta$. The dashed line shows the approximate
value of $Q$ corresponding to a squeezing factor of $\exp(-2r)=(1-\eta)/(1+\eta)$.}
\end{center}
\end{figure}

We can now determine the phase error $\delta \phi^2$ and the quantum enhancement
parameter $Q$ from the expectation values of the eight photon state according to
eq.(\ref{eq2:deltaphi}) and eq.(\ref{eq2:Q}). The result is shown in fig.
\ref{Qparameter}, together with the rough approximation of the $\eta$-dependence of
squeezing given by eq.(\ref{eq4:reta}). The validity of this approximation
is restricted to very low values of $\eta$, since the assumption that the average
photon number in mode $\hat{b}$ is much smaller than $N$ breaks down very quickly
at only eight photons. In fact, the approximation bends upward at all values of $\eta$, while the actual
quantum enhancement curve bends downward. As a result,
the values of $\eta$ necessary to achieve a given
squeezing factor are actually
higher than those expected from eq.(\ref{eq4:reta}).
For example, a squeezing factor of $1/2$ corresponding to $Q=1/3$ is only reached at
$\eta=0.427$ instead of $\eta=1/3$.
On the other hand, the approximate estimate of the maximal value of $Q=1/2$ given
by eq.(\ref{eq4:Qhalf}) is confirmed by the maximum of $Q=0.51$ at $\eta=0.85$.
At eight photons, this corresponds to an optimal squeezing factor of
0.346 times SQL, only slightly better than the rough estimate of $1/N^{3/2}=0.354$
corresponding to $Q=1/2$.

Since eq.(\ref{eq6:8state}) defines the values of the normalization coefficient
$C_8$, it is also possible to determine the improvements of coincidence
rates compared to the generation of an eight photon pair state $\mid 4;4 \rangle$
more precisely.
For example, the state at $\eta=0.427$, which has a phase error equal to 1/2 of the
SQL, has a normalization factor of $|C_8|^2=0.922$. From
eq.(\ref{eq5:PsqPpair}), we can then determine that the coincidence rate
will be $4.1 \times 10^5$ times higher than the coincidence rate of pure
down-conversion. At the optimal squeezing level of $0.346$ times SQL,
we have $\eta=0.85$ and $|C_8|^2=0.712$. According to eq.(\ref{eq5:PsqPpair}),
the coincidence rate is then about 2200 times higher than that of pure
down-conversion. It should be noted that this result is significantly higher
than the factor of 680 estimated for $\eta=1$ in
table \ref{countrate}. In general, the improvements in coincidence count
rates are quite significant and should be an important help in the
realization of non-classical multi-photon states.

\begin{figure}[t]
\begin{center}
\includegraphics[scale=1.0]{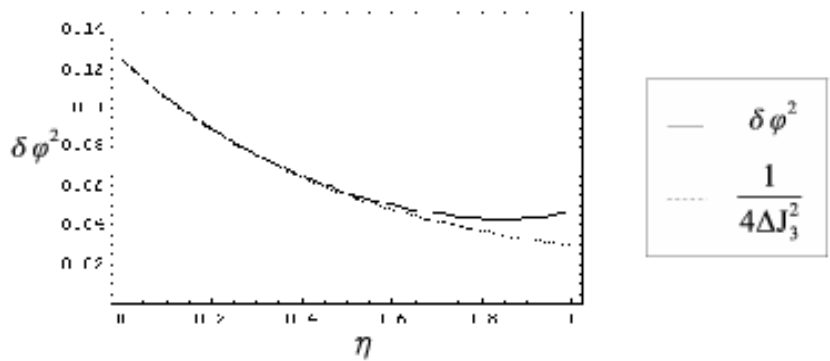}
\vspace{-0.3cm}
\caption{\label{compare}
Comparison of the phase error $\delta \phi^2$ and the optimal phase sensitivity
of $1/(4 \Delta J_3^2)$ for the eight photon state. Up to $\eta=1/2$, the state is
an optimally squeezed state. Above $\eta=1/2$, $\delta \phi^2$ reaches a minimum
value of $0.043$ at $\eta=0.845$, while $1/(4 \Delta J_3^2)$ continues to drop
towards the HL of $0.0156$.}
\end{center}
\end{figure}

Next, we take a closer look at the kind of state described by the eight
photon output given in eq.(\ref{eq6:8state}). In section \ref{sec:operators},
we have shown that the state is approximately equal to a minimal
uncertainty state of the Mandelstam-Tamm inequality (\ref{eq2:minideltaphi})
for the $\hat{J}_2$ estimator, so that the phase error defined by
eq.(\ref{eq2:deltaphi}) should be equal to the optimal phase sensitivity of
$1/(4 \Delta J_3^2)$. As a test of this approximation, we can now compare the
actual values of the phase error $\delta \phi^2$ and the optimal phase
sensitivity of $1/(4 \Delta J_3^2)$ obtained from the expectation values of the
actual eight photon state. The result of this comparison is shown in fig.
\ref{compare}. Interestingly, the phase error $\delta \phi^2$ is indeed close
to its optimal value of $1/(4 \Delta J_3^2)$ up to about $\eta=0.5$. This means
that the state is very close to a minimal uncertainty state defined according to
eq.(\ref{eq2:optimalcondition}), even though the relation between $\exp(-2r)$
and $\eta$ given by eq.(\ref{eq4:reta}) seems to break down at much lower values
of $\eta$.

At $\eta>1/2$, the optimal phase sensitivity of $1/(4 \Delta J_3^2)$
continues to drop, but the phase error $\delta \phi^2$ for the $\hat{J}_2$ estimator
levels off and reaches a minimum of about 0.043 at $\eta=0.845$. At higher $\eta$, the
$\hat{J}_2$ based estimate actually becomes worse, even though the potential phase
sensitivity indicated by the generator uncertainty $\Delta J_3^2$ continues to improve.
This means that the state at $\eta>1/2$ requires a different estimation procedure
to achieve its optimal phase sensitivity of $1/(4 \Delta J_3^2)$
\cite{Pez07}. As we have shown
elsewhere, the state eventually evolves into a state that is very close
to the maximally path entangled state at $\eta=2$
\cite{Hof07}. Above $\eta=0.5$, the eight
photon state thus makes a transition from a squeezed state with an optimal
phase estimator of $\hat{J}_2$ to a superposition state with an optimal phase
estimator that must take into account multi-photon coherences between the paths.

\begin{figure}[t]
\begin{center}
\includegraphics[scale=0.8]{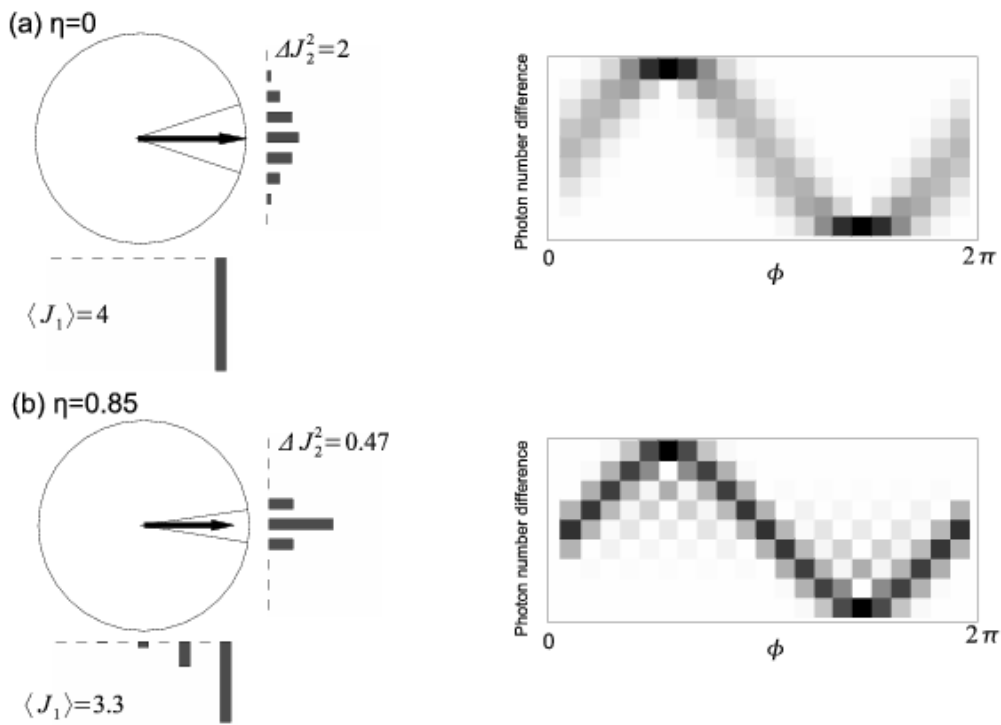}
\caption{\label{joutphi}
Illustration of the quantum statistics of eight photon states for (a) the unsqueezed state
with $\eta=0$ and (b) the highly squeezed state at $\eta=0.85$. The graphs on the left
show the $\hat{J}_1$ and $\hat{J}_2$ distributions of the input and the output at $\phi=0$.
The graphs on the right show the phase dependence of the probability distributions as
a contour plot. The high probability regions indicated by the darker shadings reproduce
the classical sinusoidal dependence of the output photon number difference
$\hat{J}_2(\phi)$.}
\end{center}
\end{figure}

We can thus see that the quantum interferometric squeezed state naturally combines
aspects of multi-photon interference fringes with the classical dependence of the
output photon number difference $\hat{J}_2$ on the phase shift $\phi$ in the
interferometer. Since the relation between the measurement probabilities themselves
and the $\phi$ dependence of $\hat{J}_2$ given by eq.(\ref{eq2:Jphi}) may not be
immediately apparent, we conclude our numerical analysis with an illustration of the
phase dependence of the complete output photon number distribution. Fig. \ref{joutphi}
(a) shows the photon statistics of the unsqueezed state at $\eta=0$, and fig.
\ref{joutphi} (b) shows the corresponding statistics at $\eta=0.85$, close to the
maximally squeezed state. The graphs on the left hand side show the $\hat{J}_1$
distribution and the $\hat{J}_2$ distribution, while the graphs on the right handside
show a contour plot of the phase dependence of the output photon number distributions
defined by the eigenstates of $\hat{J}_2(\phi)$ given by eq.(\ref{eq2:Jphi}). The shading
along each line represents the probability of a specific measurement outcome. An artificial
discretization of phase into 20 intervals of $\pi/10$ has been used to simplify the
plot. We can see from the graphs on the right hand side that the peak of the probability
distribution follows the sine-pattern expected from classical interference. However,
the unsqueezed state (a) is rather broad at $\phi=0, \pi, 2 \pi$ and sharp at $\phi=\pi/2$
and $3 \pi/2$, while the squeezed state (b) is sharper at $\phi=0, \pi, 2 \pi$ and shows
periodic oscillations within a broadened $\hat{J}_2(\phi)$ distribution at $\phi=\pi/2$
and $3 \pi/2$. It is thus possible to visualize both the classical aspects of field
amplitude interference and the quantum aspects of probability amplitude interferences
in the same graph by plotting the phase dependent many photon distribution of the
quantum state.

\section{Conclusions}
\label{sec:conclusions}
\vspace{-0.2cm}

In conclusion, we have shown how the interferometric addition of down-converted
photon pairs to weak coherent light can produce $N$-photon squeezed
states that are minimum uncertainty states of the standard phase estimation
based on the output photon number difference $\hat{J}_2$.
Significantly, the amount of squeezing depends only on the ratio of $N$ times
the down-conversion amplitude $\gamma$ and the squared coherent amplitude
$\alpha$. In typical down-conversion experiments using
coincidence counting, $N/\alpha^2 \gg 1$ allows an
enhancement of the negligibly small downconversion
amplitude $\gamma \ll 1$ to the very high squeezing levels
described by $\eta$ close to one.
It is therefore possible to obtain very high squeezing
levels, even at very low down-conversion rates.

In our derivations,
we have tried to emphasize the rather intriguing relation between measurement
noise and quantum interference in Hilbert space. It is thus possible to directly
understand squeezing as a destructive interference between the quantum noise
terms originating from the laser light and from the down-conversion, as discussed
in section \ref{sec:qi}. However, the same effect can also be represented in
terms of an operator relation that defines the minimal uncertainty of the
Mandelstam-Tamm uncertainty relation, as explained in section \ref{sec:MZ}.
Since we can derive the general operator relation (\ref{eq4:generaleta})
for the $N$ photon state $\mid \eta \rangle$ generated by interference of
laser light and down-converted light, it is possible to identify the squeezing
effect even more directly based on an approximate version of this operator
relation, as shows in section \ref{sec:operators}.

Our results indicate that we can obtain any squeezing level between the SQL and
the geometric mean of the SQL and the HL by varying the ratio of the
down-conversion amplitude and the squared coherent light amplitude.
For a given level of photon pair emission in the down-conversion, any desired
squeezing can be obtained by varying the laser amplitude.
As explained in section
\ref{sec:state}, the coincidence count rates for these phase sensitive $N$-photon
states can then be orders of magnitude higher than those observed for pure
down-conversion. In particular, there is a trade-off between the coincidence count
rates and the amount of squeezing that can be adjusted to fit the experimental
possibilities. The generation of $N$-photon squeezed states by interfering coherent light
and down-converted light at a beam splitter should thus be ideally suited for
beating the SQL of quantum metrology at increasing photon numbers $N$.

On a more fundamental level, the present approach should also help to improve
our understanding of quantum metrology by bridging the gap between the $N$
photon interference effects of path entangled states and the continuous
improvement of phase sensitivity by squeezing. We therefore hope that the present
work will open up new frontiers in the study of quantum phase measurements.

\section*{Acknowledgment}
Part of this work has been supported by the Grant-in-Aid program
of the Japanese Society for the Advancement of Science and by the
JST-CREST project on quantum information processing.


\vspace{0.5cm}

\begin{thebibliography}{xyz00}


\bibitem{Gio06}
Giovannetti V, Lloyd S and Maccone L 2006 \PRL {\bf 96} 010401


\bibitem{Eda02}
Edamatsu K, Shimizu R and Itoh T 2002 \PRL {\bf 89} 213601

\bibitem{Kok02}
Kok P, Lee H and Dowling J P 2002
\PR A {\bf 65} 052104

\bibitem{Fiu02}
Fiurasek J 2002 \PR A {\bf 65} 053818

\bibitem{Zou02}
Zou X, Pahlke K and Mathis W 2002 \PR A {\bf 66} 014102

\bibitem{Pry03}
Pryde G J and White A G 2003 \PR A {\bf 68} 052315

\bibitem{Hof04}
Hofmann H F 2004 \PR A {\bf 70} 023812

\bibitem{Wal04}
Walther P, Pan J W, Aspelmeyer M, Ursin R, Gasparoni S
and Zeilinger A 2004 Nature (London) {\bf 429} 158

\bibitem{Mit04}
Mitchell M W, Lundeen J S and Steinberg A M 2004
Nature (London) {\bf 429} 161

\bibitem{Sha04}
Shafiei F, Srinivasan P and Ou Z Y 2004
\PR A {\bf 70} 043803

\bibitem{Eis05}
Eisenberg H S, Hodelin J F, Khoury G and Bouwmeester D 2005 \PRL {\bf 94} 090502

\bibitem{Liu06}
Liu B and Ou Z Y 2006
\PR A {\bf 74} 035802

\bibitem{Nie07}
Nielsen A E B and Molmer K 2007 \PR A {\bf 75} 063803

\bibitem{Cab07}
Cable H and Dowling J P 2007 \PRL {\bf 99} 163604


\bibitem{Res07}
Resch K J, Pregnell K L, Prevedel R, Gilchrist A,
Pryde G J,O'Brien J L and White A G 2007
\PRL {\bf 98} 223601


\bibitem{Hol93}
Holland M J and Burnett K 1993 \PRL {\bf 71} 1355

\bibitem{Ste02}
Steuernagel O 2002 \PR A {\bf 65} 033820

\bibitem{Wan05}
Wang H and Kobayashi T 2005 \PR A {\bf 71} 021802(R)


\bibitem{Hof06}
Hofmann H F 2006 \PR A {\bf 74} 013808

\bibitem{Sun06}
Sun F W, Liu B H, Huang Y F, Ou Z Y and Guo G C 2006
\PR A {\bf 74} 033812


\bibitem{Nag07}
Nagata T, Okamoto R, O`Brien J, Sasaki K and Takeuchi S 2007
Science {\bf 316} 726


\bibitem{Sun07}
Sun F W, Liu B H, Gong Y X, Huang Y F, Ou Z Y and Guo G C
2007 {\it Preprint} arXiv:0710.2922v1

\bibitem{Cav81}
Caves C M 1981 \PR D {\bf 23} 1693

\bibitem{Yur86}
Yurke B, McCall S L and Klauder J R 1986 \PR A {\bf 33} 4033

\bibitem{Kit93}
Kitagawa M and Ueda M 1993 \PR A {\bf 47}, 5138

\bibitem{Hil93}
Hillery M and Mlodinow L 1993 \PR A {\bf 48} 1548

\bibitem{Win94}
Wineland D J, Bollinger J J, Itano  W M
and Heinzen D J 1994 \PR A {\bf 50} 67

\bibitem{Par95}
Paris M G A 1995 \PL A {\bf 201} 132

\bibitem{Ber00}
Berry D W and Wiseman H M 2000 \PRL {\bf 85} 5098

\bibitem{Com05}
Combes J and Wiseman H M 2005 J. Opt. B: Quantum and Semiclass. Opt. {\bf 7} 14

\bibitem{Lu02}
Lu Y J and Ou Z Y 2001 \PRL {\bf 88} 023601


\bibitem{Bra96}
Braunstein S L, Caves C M and Milburn G J
1996 {\it Ann. Phys.} {\bf 247} 135

\bibitem{Bra94}
Braunstein S L and Caves C M 1994 \PRL {\bf 72} 3439

\bibitem{Hof07}
Hofmann H F and Ono T 2007 \PR A {\bf 76} 031806(R)

\bibitem{Pez07}
Pezze L and Smerzi A 2007 {\it Preprint} arXiv:0705.4631v1

\end{thebibliography}
\end{document}